\documentclass[nofootinbib,floats,floatfix,eqsecnum,prd,aps]{revtex4}
\usepackage{dcolumn,epsfig,minitoc}
\usepackage{amssymb,amsmath}
\usepackage[usenames]{color}

\def\a{\alpha}\def\b{\beta}\def\d{\delta}
\def\f{\phi}\def\g{\gamma}
\def\l{\lambda}\def\m{\mu}\def\n{\nu}\def
\p{\pi}

\def\D{\Delta}\def\L{\Lambda}
\def\O{\Omega}
\def\OM{dx_{i}dx_{i}¥}
\def\l{\lambda}
\def\L{\Lambda}

\def\inf{\infty}\def\id{\equiv}\def\mo{{-1}}

\def\coo{coordinates }

\def\bb{black brane }\def\DW{domain wall }
\def\o1{$O_{1}$}
\def\o2{$O_{2}$}

\def\lb{\label}
\newcommand{\beq}{\begin{equation}}
\newcommand{\eeq}{\end{equation}}
\newcommand{\bea}{\begin{eqnarray}}
\newcommand{\eea}{\end{eqnarray}}

\begin{document}
\title{ PHASE TRANSITION AND  HYPERSCALING VIOLATION FOR  SCALAR BLACK BRANES}
\author{Mariano Cadoni}
\affiliation{Dipartimento di Fisica, Universit\`a di Cagliari and INFN, Sezione di
Cagliari - Cittadella Universitaria, 09042 Monserrato, Italy. }
\author{Salvatore Mignemi}
\affiliation{Dipartimento di Matematica, Universit\`a di Cagliari and INFN, Sezione di
Cagliari - viale Merello 92, 09123 Cagliari, Italy. }

\date{\today}



\begin{abstract}
We  investigate the thermodynamical behavior and
the scaling symmetries of the scalar dressed black brane (BB)
solutions  of a recently
proposed, exactly integrable  Einstein-scalar
gravity  model \cite{Cadoni:2011yj}, which also arises as 
compactification of $(p-1)$-branes with a smeared charge.
The extremal, zero temperature,
solution is a scalar soliton
interpolating between a
conformal  invariant AdS vacuum in the near-horizon region
and a scale covariant  metric
(generating  hyperscaling violation on the boundary field theory)
asymptotically. We show
explicitly  that for the boundary field theory this implies
the emergence of an UV
length scale (related to the size of the brane),
which decouples in the IR, where conformal invariance is
restored. We also show that at high temperatures the system undergoes
a phase transition. Whereas at small temperature
the Schwarzschild-AdS BB is stable, above a critical temperature
the scale covariant, scalar-dressed BB solution,
becomes energetically preferred. We  calculate the critical
exponent $z$ and the hyperscaling violation parameter
$\theta$ of the scalar-dressed phase.
In particular we show that $\theta$ is always negative.
We also show that the above features are not a peculiarity of the
exact integrable model of Ref.\ \cite{Cadoni:2011yj}, but  are a quite generic
feature of  Einstein-scalar and Einstein-Maxwell-scalar gravity models
for which the squared-mass of the scalar field $\phi$ is
positive and the potential vanishes exponentially as $\phi\to- \infty$.

\end{abstract}

\maketitle




\section{Introduction}
In recent years holographic methods have been widely used to
investigate features of strongly interacting quantum field theories
(QFT),  in particular for what concerns possible applications to
condensed matter systems \cite{Hartnoll:2008vx,Hartnoll:2008kx,
Horowitz:2008bn,Herzog:2009xv,Hartnoll:2009sz,Charmousis:2009xr,
Cadoni:2009xm,Goldstein:2009cv, Gouteraux:2011ce,Charmousis:2010zz}.

 These holographic methods have been  developed following  rather 
 closely the anti-de Sitter/conformal field theory (AdS/CFT)
correspondence paradigm. This means that the focus has been  mainly on
bulk AdS gravity in $d+2$ dimensions and its $(d+1)$-dimensional
conformal  field theory  duals. In fact, it has turned out that simply  putting
in the AdS background a black hole (black brane), a
non trivial scalar field configurations  and eventually 
finite electromagnetic charge density, a very rich phenomenology  in
the dual QFT
could be obtained. This includes spontaneous  breaking of the
$U(1)$ symmetry,  phase transitions triggered by scalar condensates
and non-trivial transport properties of the dual field theory
\cite{Hartnoll:2008vx,Hartnoll:2008kx,
Horowitz:2008bn,Herzog:2009xv,Hartnoll:2009sz,Charmousis:2009xr,
Cadoni:2009xm,Goldstein:2009cv, Gouteraux:2011ce,Charmousis:2010zz,Bertoldi:2010ca,
Cadoni:2011kv,Lee:2011zzf,Fujita:2012fp}.

An interesting and welcome byproduct of these investigations has
been the realization of the importance played by non-AdS gravitational
backgrounds and the related  dual nonconformal QFTs.
When the self-interaction potential and the Maxwell tensor/scalar 
field
coupling function  behave exponentially, a quite generic
feature of the  bulk  Einstein-Maxwell-scalar  gravity theory
is the emergence in the extremal, near-horizon  (infrared) region
of  solutions breaking the full conformal symmetry of the
(ultraviolet) AdS vacuum, but  still preserving some symmetry of the 
AdS
background\footnote{ Throughout this paper
we use infrared (IR) and ultraviolet (UV) to denote,
respectively, the infrared and ultraviolet region of the dual
boundary field
theory. In terms of the bulk gravity theory they correspond
respectively to the near-horizon  and asymptotic
region.}
\cite{Goldstein:2009cv,Cadoni:2009xm,Goldstein:2010aw,Charmousis:2010zz,
Cadoni:2011kv,Bertoldi:2011zr,Iizuka:2011hg,Kim:2012nb,Dong:2012se,
Wu:2012fk}.

The existence of these non-AdS solution is not just a peculiarity
of theories with non minimal (exponential) couplings between the scalar and the
Maxwell field
\cite{Goldstein:2009cv,Charmousis:2009xr,Cadoni:2009xm,Cai:2004iy,
Charmousis:2010zz,Li:2012ib,Gouteraux:2011qh,Iizuka:2012wt}. They  also arise
as solutions of string theory
and supergravity constructions
\cite{Perlmutter:2010qu,Narayan:2012hk,Dong:2012se,Dey:2012rs}.
From a purely gravitational point of view their existence
is   a generic  consequence of
the  presence of a non-trivial scalar with  a self interaction
potential behaving exponentially \cite{Cadoni:2011nq}.

According to the kind of symmetry that is preserved in the IR, these
non-AdS  metrics  can be classified in two classes.
To the first
class, called Lifshitz,   belong metrics for which a scale isometry,
under which timelike and spacelike coordinates transform with a
different exponent,
is preserved but the conformal and Poincar\'e isometry of the
$d+1$-dimensional spacetime is broken \cite{Dehghani:2011tx,Kachru:2008yh,Bertoldi:2009dt,
Bertoldi:2009vn,Bertoldi:2011zr,
Charmousis:2010zz,Gouteraux:2011ce}. They could be relevant
for the holographic description of quantum phase transitions.
In analogy with critical systems in condensed matter physics, such
metrics are characterized by a
dynamical critical exponent $z$ describing the anisotropy of scaling
of the space and time coordinates.
The second class of these non-AdS metrics is characterized by
breaking of the scale isometry, but  the $(d+1)$-dimensional Poincar\'e invariance is
preserved.
In the literature they are often referred to as domain
wall (DW) solutions
\cite{Kaitscheider:2009as,Boonstra:1998mp,Perlmutter:2010qu,Cadoni:2011nq,Cadoni:2011yj},
in analogy with the DW solutions of supergravity (SUGRA) theories.

Very recently, it has been realized that these two classes of metrics
are particular cases of a more general class of metrics that are
not scale invariant (but only scale covariant) 
\cite{Gouteraux:2011ce,Cadoni:2011nq,Cadoni:2011kv} and  lead to
hyperscaling violation in the dual field theory
\cite{Huijse:2011ef,Dong:2012se,Kim:2012nb}.
They are
characterized by  two parameters, the critical exponent $z$ and the
hyperscaling violation parameter $\theta$, which gives the
transformation weight of the infinitesimal length $ds$ under scale
transformations and the corresponding scaling behavior of the free
energy as function of the temperature \cite{Huijse:2011ef,Dong:2012se}. Scale covariant
metrics are a very promising
framework for the holographic description of hyperscaling violation
in condensed matter critical systems  (e.g Ising models) \cite{Fisher:1986zz}.

The crucial  holographic feature of the class of models
allowing for   scale-covariant
metrics,   is the emergence of a length scale in the IR \cite{Dong:2012se}.
If the theory has an UV fixed point, this length scale 
decouples in the UV because of the conformal invariance of the AdS
background.  The emergence of a  length scale in the IR has  several
interesting physical consequences, which have been used for the
description of Fermi surfaces and for the related area-law
violation of entanglement entropy
\cite{Dong:2012se,Huijse:2011ef,Shaghoulian:2011aa,Ogawa:2011bz}.

Until now the standard setup used for obtaining, dynamically,
scale -covariant
metrics, is given by Einstein-scalar gravity, possibly coupled
-- minimally or non-minimally  -- with a $U(1)$ field. The
self-interaction potential $V(\phi)$ for the scalar field $\phi$
is assumed to have a negative local maximum at $\phi=0$, with a
corresponding scalar tachyonic excitation whose  mass is slightly above
the  Breitlohner-Freedman (BF) bound. The $\phi=0$ solution  of the
gravity theory
corresponds to the  AdS vacuum, whereas under suitable conditions
-- typically an exponentially behavior of the potential and/or 
scalar/Maxwell tensor coupling
functions -- the theory admits  black brane solutions with scalar hair
that in the near-extremal regime approach the scale covariant
metric.

The requirement that the $\phi=0$ solution be a maximum of $V(\phi)$
is necessary  for the existence of non trivial black brane solutions
with AdS asymptotics. In fact, usual no-hair theorems based on the
positive energy theorem forbid the existence of black brane (BB) solutions with
AdS asymptotics  when the squared-mass of the scalar is positive
\cite{Townsend:1984iu,Torii:2001pg,Hertog:2006rr}.
However, in a recent paper it has been shown that these no-hair
theorems can be circumvented by giving up the condition that  the BB
solution has AdS asymptotics \cite{Cadoni:2011nq,Cadoni:2011yj}.
In Ref.\ \cite{Cadoni:2011yj} we have
derived  exact hairy, asymptotically non-AdS, black brane
solutions of a
Einstein-scalar gravity  model with positive squared  mass for the
scalar field.
We have shown that the  extremal limit of these BB
solutions  is a scalar soliton which interpolates  between an AdS
vacuum in the near-horizon region and a scale covariant, DW, solution in the
asymptotic region. Moreover, these BB solutions arise as Kaluza-Klein 
compactification of black $2$-branes with a smeared charge supported by 
a $4$-form field strength \cite{Gouteraux:2011ce}.

The results of Refs.\ \cite{Cadoni:2011yj,Gouteraux:2011ce} open up 
the possibility of realizing
an alternative
scenario in which the usual roles played by the IR and  UV regions are
reversed. In the dual QFT, the IR physics is determined by an infrared
fixed point, i.e.\ by  the
conformal symmetry of the AdS vacuum, whereas the UV behavior is
characterized by hyperscaling violation and by an emergent UV length
scale. In this paper we
study in detail this alternative framework. Building on
the results of Ref.\ \cite{Cadoni:2011yj}, we first show that the scalar soliton
obtained  as  the zero-temperature extremal limit of the BB solutions
of Ref.\ \cite{Cadoni:2011yj}
gives a nice realization of this scenario (Section \ref{sect:BB}).
Using the Euclidean action formalism  we  derive  a consistent
formulation for the thermodynamics of our BB solutions (Section
\ref{ther}).  This allows us to show that in these Einstein-scalar
gravity models the Schwarzschild-anti de Sitter (SAdS) solution is
unstable above a critical temperature $T_{c}$. At small
temperatures the SAdS solution with constant vanishing scalar field
is energetically preferred with respect to  a  scalar-dressed, BB solution
with DW asymptotics. This is not the case for
$T>T_{c}$, and the thermodynamical system undergoes a first-order
phase transition SAdS $\to$ scalar BB solution (Section \ref{sect:pt}).
The thermodynamical behavior of the solutions, the emergence of a length
scale in the UV and its decoupling in the IR
 have a natural explanation in terms of the symmetries of the
solutions.
In particular, we show that the asymptotic  solution is a
scale-covariant  metric characterized by critical exponent
$z=1$ and {\sl negative} hyperscaling-violating parameter $\theta$
(Section \ref{sect:hyper}).
Although most of the calculations have been performed for the exact
BB solution of the  Einstein-scalar gravity model introduced in  Ref.\
\cite{Cadoni:2011yj}, we show that the most important features (SAdS $\to$ scalar
BB phase transition, hyperscaling violation)  are quite generic
features of a broad class of Einstein-scalar and
Einstein-Maxwell-scalar gravity models.  Sufficient conditions for
their presence are the existence of a negative minimum of the potential $V(\phi)$
and an exponential behavior, $V(\phi)\sim e^{\alpha \phi}, \, \a>0$, for
$\phi\to- \infty$ (Section \ref{sect:gm}).

\section{Black brane solutions}\lb{sect:BB}
We consider  static, radially symmetric, planar solutions of Einstein gravity
minimally coupled  to a scalar field with self-interaction potential
$V(\phi)$.
The  action is
\beq\lb{action}
I=\frac{1}{16\pi G}\int d^{4}x \left[R-
2(\partial \phi)^{2} -V(\phi)\right].
\eeq
where $\phi$ is a scalar field. Following widespread conventions
we  set $G=1/(16\pi)$.

We will consider models with a potential $V(\phi)$ satisfying the
following conditions: 1) $V(\phi)$ has a local minimum for $\phi=0$
with $V(0)<0$; 2) The potential approaches zero exponentially as 
$\phi\to-\infty$.
We will therefore assume the following behavior,
\beq\lb{b1}
V(\phi)\sim e^{2h\phi},\, h>0,\, {\rm for}\,\, \phi\to-\infty;\quad
V(0)=-\frac{6}{L^{2}},\ V'(0)=0,\ V''(0)>0,
\eeq
where $L$ is the AdS length. The previous conditions ensure the existence of
an AdS$_{4}$ vacuum   and of a Schwarzschild-AdS
(SAdS)  black brane solution with $\phi=0$.  On the other hand  standard
positive energy theorems  forbid the existence of  black brane
solutions with scalar hair and AdS asymptotics \cite{Townsend:1984iu,Torii:2001pg,Hertog:2006rr}.

We first focus on a particular model, which is exactly
integrable. In section \ref{sect:gm} we will extend our
considerations to general models satisfying (\ref{b1}).
The particular model we consider is characterized by the potential
\beq\lb{k3}
V(\phi)= -\frac{6}{\g L^{2}}\left(e^{2\sqrt3\beta
\phi}-\beta^{2}e^{\frac{2\sqrt3}{\beta}\phi}\right), \qquad
\g=1-\b^2,\eeq
where $\b$ is a real parameter. The limit $\b\to1$ gives a Liouville (purely exponential) potential,
whereas for $\b=0$ the potential reduces to a  cosmological constant.
The action is invariant under the duality transformation $\b\to1/\b$.

The gravity-scalar model defined by the potential (\ref{k3})
has several interesting features.  It is a fake SUGRA model, i.e.\
the potential can be derived from a superpotential $P(\phi)=(1/\g L)
\left(e^{\sqrt3\beta\phi}-\beta^{2}e^{\frac{\sqrt3}{\beta}\phi}\right)$, see Ref.\
\cite{Cadoni:2011yj}.
The model is exactly integrable, since the field equations for static, radially symmetric,
planar solutions  can be reduced to that of a  Toda molecule.
The model allows to circumvent  standard no-hair theorems -- it admits
black brane solutions with non-AdS asymptotics.
Last but not least, the model (\ref{k3}) arises as Kaluza-Klein 
compactification on a $q$-dimensional compact space  
$\bold{K}^{q}$ of a  black $(p-1)$-brane  with a smeared charge, which  
is 
solution of the action \cite{Gouteraux:2011ce},
\beq
S= \int d^{p+q+1}x\, \sqrt{-g}\left( R- 
\frac{1}{2(n+2)!}G_{(n+2)}\right),
\eeq
where $G_{(n+2)}$ is the field-strength form.
The action (\ref{action})  with the potential (\ref{k3}) is obtained 
by Kalunza-Klein compactification of $\bold{K}^{q}$  
considering   $p=3$, i.e a $2$-brane,  with $n=p-1=2$, so that  
$G_{(n+2)}$ can be dualized to a scalar \cite{Gouteraux:2011ce}.

\begin{figure}[ht]
\begin{center}
\begin{tabular}{cc}
\epsfig{file=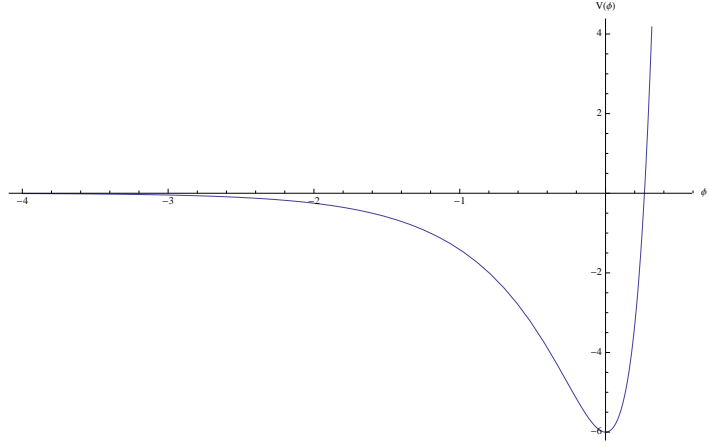,width=9.3cm,angle=0}
\end{tabular}
\caption{Plot of the potential $V(\phi)$ for a selected value of the
parameter $\b=1/2$ and $L=1$.
\label{fig:potential}}
\end{center}
\end{figure}

The potential (\ref{b1}) has a minimum at $\phi=0$ with $V(0)=-6/L^{2}$,
corresponding to an AdS$_{4}$ vacuum and related local scalar
excitation of positive squared-mass  $m^{2}=18/L^{2}$.
For $\phi\to -\infty$ the potential approaches zero, while it diverges
for $\phi\to\infty$. The plot  of $V(\phi)$ for a selected value
of the parameter, $\b=1/2$ and $L=1$, is shown in Fig.\ \ref{fig:potential}.

It has been shown in \cite{Cadoni:2011yj} that the present model admits 
two classes of static
black brane solutions: the SAdS black brane with constant vanishing 
scalar field and a scalar black
brane  endowed with a non trivial, $r$-dependent, scalar field profile.

The SAdS solution always exists for every value of $\b$. The
scalar field is  $\phi=0$ (minimum of
the potential  $V$) and the metric part of the solution reads
\beq\lb{SADS}
ds^{2}= - f(r) dt^{2}+f^{-1}(r) dr^{2}+r^{2}\OM, \qquad f(r)=
\frac{r^{2}}{L^{2}}- \frac{2M}{r}.
\eeq
where  $M$ is the \bb mass.

We now investigate the solution with non-trivial scalar field, starting
from the values $\beta^{2}<1$. The case $\beta^{2}>1$ will be discussed
in Sect.\ \ref{sect:bmuno}.

For $\beta^{2}<1$, the black brane solution is given by\footnote{The 
solution in this form, apart from some trivial change of notation, 
has been derived in Ref. \cite{Cadoni:2011yj}. The same solution, 
albeit in a different form, is a particular case ($p=3,n=2$) 
of the compactified black $(p-1)$-brane  solutions with a smeared charge 
derived in Ref. \cite{Gouteraux:2011ce}.}
\bea\lb{met1}
ds^2=&&\Delta(r)^{\frac{2\b^2}{3\g}}\left[\left(\frac{r}{r_0}\right)^{\frac{2}
{1+3\b^2}}\left( -\Gamma(r) dt^2+\OM\right)+
E\,\frac{\Delta(r)^{\frac{4\b^2}{3\g}}}{\Gamma(r)}\left(\frac{r}{r_{0}}\right)^{-\frac{2}
{1+3\b^2}}\,dr^2 \right],\cr
e^{2\f}=&&\left[\frac{A}{\Delta(r)}\left(\frac{r}{r_0}\right)
^{-\frac{3\g}{1+3\b^2}}\right]^{\frac{2\b}{\sqrt3\g}},
\eea
where
$\Gamma(r)= 1-{\m_1}\left(\frac{r_0}{r}\right)^\d$,
$\Delta(r)= 1+{\m_2}\left(\frac{r_0}{r}\right)^\d$, $\d=3\g/(1+3\b^{2})\,$,
$A= \sqrt{\m_{2}(\m_{1}+\m_{2})}$,
$E=\left(\frac{\g L}{(1+3\b^{2})r_0}\right)^{2}A^{-2\b^2/\g}$,
$\m_{1,2}$ are dimensionless free parameters and
$r_0$ is  a length scale  that
must be introduced in order to get the correct physical dimensions.
Physically, $r_{0}$ is related to the size of the brane. In fact, it
drops out completely from the solution (\ref{met1}) just by rescaling
appropriately the coordinate $x_{i}$ (and the time $t$) and by
redefining the parameters $\m_{1,2}$.
In principle  $r_{0}$ can take any value, but
for a holographic interpretation to hold, we require $r_0\gg L$.
The  parameters $\m_1\ge 0$ and $\m_{2}\ge 0$ are  assumed to be
small numbers, even if in principle $\m_1$ can have a greater scale than $\m_2$.
Note that the metric functions $g_{tt}$ and $g_{ii}$ can be arbitrarily rescaled,
since no natural normalization exists for the \coo $t$ and $x_i$.

The  asymptotic behavior of the solution (\ref{met1}) for $r\to\infty$ is
that of a domain wall: $ds^{2}= (r/r_0)^{\eta}\left( -dt^{2}+\OM\right)+
(r/r_0)^{-\eta}dr^{2},$ where $\eta=2/(1+3\b^2)$, and the scalar field behaves
logarithmically, $\phi= -[(\sqrt{3}\b)/(1+3\b^{2})]\log (r/r_0)$. One may also
 define a scalar charge, which is proportional to $\m_2$.

For $\m_1>0$, the metric (\ref{met1}) exhibits a singularity at $r=0$,
shielded by a horizon at $r/r_0=\m_1^{\,1/\d}$, and therefore represents a
regular black brane. Owing to the fact that the scalar $\phi$
depends on $\m_{1}$,  the  existence of this \bb solution is perfectly
consistent with the no-hair theorem of Ref.\ \cite{Cadoni:2011nq}.
Notice that for $\m_{2}>0$, although the scalar field  remains finite at $r=0$,
the scalar curvature $R$ of spacetime diverges as $
R\sim r^{-3(1+\b^{2})/(1+3\b^{2})}$.
For $\m_{2}=0$ the metric (\ref{met1}) describes the \bb solution 
with a $\phi\sim \ln r$  short distance singularity discussed in Ref. 
\cite{Cadoni:2011nq}. 

A detailed discussion of the causal structure of the
spacetimes (\ref{met1}) is beyond the scope of this paper. We just
note that the causal structure of two-dimensional models with metric behaving as
$ds^{2}=-r^{a}dt^{2}+r^{-a}dr^{2}$ with $0<a\le 2$,  has been already
discussed in Ref. \cite{Cadoni:1995dd}. From those results it follows immediately
that  the $r=\infty$ asymptotic region is spacelike for
$\b^{2}\ge 1/3$, whereas it becomes timelike for  $\b^{2}<1/3$.
The timelike boundary at $r=\infty$, for $\b^{2}<1/3$, makes the
spacetime conformally equivalent to AdS$_{4}$, and is therefore a crucial
ingredient  for a holographic interpretation of these solutions
\cite{Kaitscheider:2009as,Boonstra:1998mp,Perlmutter:2010qu}.
Notice that in the coordinates used in Eq.\ (\ref{met1}),
$r\to\infty$ describes the UV regime of the
dual QFT, whereas $r\to 0$ corresponds to the IR regime.

The limiting case  $\b^{2}=1/3$ of the solution (\ref{met1}) is
particularly simple:
\bea\lb{met1a3}
ds^2=&&\left(1+\m_{2}\frac{r_0}{r}\right)\left[-\left(r-\m_1r_0\right)dt^2+
\frac{E \left(1+\m_{2}\frac{r_0}{r}\right)^{2}}{(r-\m_1r_0)}\,dr^2
+ r\,\OM\right],\cr
e^{2\f}=&&\frac{A}{r+\m_{2}r_0}.
\eea
Asymptotically, the two-dimensional $x_{i}=$ const.\ sections of the metric
(\ref{met1a3})  describe Rindler spacetime.

For  $\m_{2}>0$, the extremal limit $\m_1=0$ of the solution (\ref{met1})
represents a regular scalar soliton \cite{Cadoni:2011yj}, i.e.\ a soliton endowed
with a non trivial scalar profile. In
fact, not only the scalar field vanishes at $r=0$,
but also the scalar curvature of the spacetime   remains finite both
at $r=0$ and $r=\infty$. The scalar soliton does not exist for $\m_{2}=0$.
In this  case the extremal limit $\m_{1}=0$  is a spacetime
with a singularity at $r=0$ and a scalar field behaving as
$\log r$.

In order to clarify  the role played by the two scales $L$
and $r_{0}$, let us now write explicitly the form of the solitonic
solution.  Setting $\m_{1}=0$, the metric (\ref{met1}) becomes
\bea\lb{met1a}
ds^2=&&\Delta(r)^{\frac{2\b^2}{3\g}}\left[
\left(\frac{r}{r_{0}}\right)^{\frac{2}{1+3\b^2}}\left(-dt^2+\OM\right)+
k^2\left(\frac{L}{r_{0}}\right)^2 \Delta(r)^{\frac{4\b^2}{3\g}}
 \left(\frac{r}{r_{0}}\right)^{-\frac{2}{1+3\b^2}}dr^2 \right],\cr
e^{2\f}=&&\left[\frac{\m_2}{\Delta(r)}\left(\frac{r}{r_{0}}\right)^{-\frac
{3\g}{1+3\b^2}}\right]^{\frac{2\b}{\sqrt3\,\g}}
\eea
where
$k=\g/(1+3\b^{2})\m_2^{-\b^2/\g} .$

In the IR limit, $r\to 0$, the scalar field approaches to zero, the length scale $r_{0}$ decouples and
the metric (\ref{met1a}) becomes,
after suitable
rescaling of $t$ and $x_i$, the metric of AdS$_{4}$, with AdS length $L$:
\beq\lb{ADS}
ds^{2}= \frac{ r^{2}}{L^{2}}\left( -dt^{2}+\OM\right)+
\frac{L^{2}}{ r^{2}}\,d r^2.
\eeq
Note that, because of scale invariance, $g_{ r
r}$ is independent from the scale of $ r$.
In this regime, that holds for $r\lessapprox L$, conformal invariance is restored.

Conversely, in the UV limit $r\to\infty$, it is the AdS length $L$ that
decouples, and the metric (\ref{met1}) can be written entirely
in terms of $r_{0}$ only. In fact taking in Eq. (\ref{met1a}) the
limit $r\to\infty$ one gets, after suitable rescaling of $r,t,x_{i}$,
the scale-covariant \DW metric
\beq\lb{DW}
ds^{2}= \left(\frac{r}{r_{0}}\right)^{\frac{2}{1+3\b^2}}\left[ -dt^{2}+\OM\right]+
\left(\frac{r}{r_{0}}\right)^{-\frac{2}{1+3\b^2}}dr^{2},
\eeq
whereas  the scalar field behaves  logarithmically
$\phi= -[(\sqrt{3}\b)/(1+3\b^{2})]\log (r/r_{0})$.
This regime is attained for $r\gg r_0$. In this case, the form of $g_{rr}$
depends on the rescaling or $r$.
The metric (\ref{DW}) is not invariant under scale transformations
but  transforms with a definite weight.

The extremal soliton has therefore the form of a  brane that
interpolates between a scale covariant, \DW solution in the UV  and  the AdS
spacetime  in the IR.
The metric behaves as $ds^{2}=
-(r/l_{a})^{\eta}[-dt^{2}+\OM] +(r/l_{a})^{-\eta}dr^{2}$
in the limit of small or large $r$,
with  different power $\eta$ and length-scale $l_{a}$ in the IR and UV.
This behavior has a natural interpretation in terms of the
symmetries  of the solutions (\ref{ADS}) and (\ref{DW}) as we will
explain in detail in Sect.\ \ref{sect:hyper}.

 It is also important to notice that the $r\to\infty$ UV  region of the soliton
(\ref{met1}), corresponds to $\phi\to-\infty$,  the region where
the potential $V(\phi)$ approaches zero. Conversely, the  $r=0$ IR region
 corresponds to $\phi=0$, i.e.\ to the minimum of the potential
$V(\phi)$.

It is instructive to think of  the above-described situation as the
opposite (IR exchanged with UV) of that holding in
Einstein-gravity models with holographic applications to condensed
matter systems. Typically, in the latter case one considers potentials $V(\phi)$
with a negative maximum and with the negative squared-mass $m^{2}$ of the
corresponding tachyonic excitation slightly above the BF bound.
In this case the dual field theory has an UV fixed point,
corresponding  to the AdS vacuum and the  IR  corresponds to
strong self-interaction of the scalar $V(\phi)=-\infty$.
The length-scale $r_{0}$ is an IR scale, which decouples in the UV,
where conformal invariance is restored.

Let us now come back to the general solution   (\ref{met1})  with
$\m_{1}\neq 0$. Because (\ref{met1}) is a global solution interpolating
between the IR and the UV regime and because the two regimes are characterized
by two different length  scales $r_{0}$ and $L$, it is convenient to define
dimensionless coordinates $t,r$. Moreover this will allow us to write
the solution in a simpler form.
Rescaling the coordinates as follows,
\beq\lb{j5}
\frac{r}{r_{0}}\to r D^{(1+3\b^{2})/3\g},\quad t\to \left(  \frac{\g L}
{1+3\b^2}\right) t,\quad  x_{i}\to x_{i}D^{-1/3\g},
\eeq
the solution (\ref{met1})   takes the form,
\bea\lb{met2}
ds^2=&&\left(\frac{\g L}{1+3\b^{2}}\right)^{2}\left[- \Delta^{\frac{2
\b^2}{3\g}}(r)\left(1-\frac{\n_1}{r^\d}\right)r^{\frac{2}{1+3\b^2}}dt^2+
\frac{ \Delta^{\frac{2\b^2}{\g}}(r)\,dr^2}{(1-\frac{\n_1}{r^{\d}})
r^{\frac{2}{1+3\b^2}}}\right]+
\Delta^{\frac{2\b^2}{3\g}}(r)r^{\frac{2}{1+3\b^2}}\OM,\cr
&&\cr
e^{2\f}=&&\Delta^{-\frac{2\b}{\sqrt3\,\g}}(r)r^{-\frac{2\sqrt3\,\b}{1+3\b^2}},
\eea
where now $\Delta(r)= 1+\frac{\n_2}{r^\d}$ and
$\n_{1,2}=\frac{\m_{1,2}}{A}=
\frac{\m_{1,2}}{\sqrt{\m_{2}(\m_{2}+\m_{1})}}$.

Unfortunately, the  dimensionless parameters $\n_{1}$ and $\n_{2}$
so defined are not independent, but are constrained by  the algebraic relation
\beq\lb{con}
\n_{2}(\n_{2}+\n_{1})=1,
\eeq
which can solved for $\n_{1}$:
\beq\lb{con1}
\n_{1}= \frac{1}{\n_{2}}-\n_{2}.
\eeq
Thus, in the new  coordinates  the solution has only one independent
parameter, $\n_{2}$.
This means that one cannot vary independently the \bb temperature (or
mass)  and the scalar charge. For instance, the $T=0$ scalar soliton
has necessarily $\n_{2}=1$. The range of variation of $\n_{2}$ is 
$0\le
\n_{2}\le 1$, corresponding to $\infty\ge \n_{1}\ge 0$.

Notice that the solution (\ref{met2}) and  the constraint (\ref{con})
can be directly derived from
the general solution (\ref{met1}) by imposing that the scalar field does
not depend explicitly on $\n_{1}$. This requires $A=1$, which is
equivalent to Eq.\ (\ref{con}). With this
assumption we get, after a rescaling of the time coordinate, solution
(\ref{met2}) from solution (\ref{met1}).

The two forms of the solution are completely consistent with the
no-hair theorem of Ref.\ \cite{Cadoni:2011nq}. In the form (\ref{met1}) the
solution is parametrized by two independent parameters $\m_{1,2}$ but
the scalar field depends on both of them, i.e.\ it depends on both the
\bb temperature and the scalar charge $\m_{2}$.  When we try to make
the scalar field  independent from the temperature we are forced to impose
the constraint  (\ref{con}) and the solution has only one independent
parameter. The scalar field  depends now implicitly on the temperature.
In the form  (\ref{met2}) our solution is very similar to that of
Martinez et al.\ \cite{Martinez:2004nb}. 
Indeed, both solutions fully confirm the no-hair
theorem or Ref.\ \cite{Cadoni:2011nq}.

As we shall see, the thermodynamics of the \bb can be consistently defined only for the metric
(\ref{met2}). It is an open question if to the two-parameter solution
(\ref{met1}) can be given a physical interpretation.
Its existence
follows from the fact that the angular part of the metric has no natural
normalization in planar coordinates.

\section{Thermodynamics}\lb{ther}

To set up the thermodynamics of our black brane solutions we use the
euclidean action formalism of Martinez et al.\ \cite{Martinez:2004nb} (Our
conventions in the action (\ref{action}) correspond to setting $G=1/16\pi$
and to multiplying by 4 the kinetic term for the scalar  field in the paper of
Martinez et al.).

The thermodynamical behavior of the solutions (\ref{met1})
is problematic due to the occurrence of divergences in
the boundary  action, that determines the mass of the solution.
This divergences are difficult to remove because the scalar depends
explicitly on the parameter $\m_{1}$. On the other hand, as we will show
later in this section, the formulation of the thermodynamics of the
solution (\ref{met2}) with the parameters $\n_{1,2}$ constrained
by Eq. (\ref{con}) is free from this difficulty.
It is not completely clear to us whether a
consistent thermodynamical interpretation  of the two-parameter
solution (\ref{met1}) exists. We will not address this issue here but
we will simply   formulate   the thermodynamics  of the solution written
in the form (\ref{met2}), (\ref{con}). In the calculation we will first consider
$\n_{1,2}$ as independent parameters, and then use the constraint
(\ref{con1}) at the end of the calculations.

Standard calculations give for the temperature $T$ and entropy $S$  of
our \bb solution (\ref{met2})
\beq\lb{TS}
T=\frac{1}{4\pi}\frac{3\g}{1+3\b^{2}}(\n_{1}+\n_{2})^{-\frac{2\b^{2}}
{3\g}}\n_{1}^{1/3}, \qquad S= \frac{\O R^{2}}{4G^2}= 4\pi \O
(\n_{1}+\n_{2})^{\frac{2\b^{2}}{3\g}}\n_{1}^{2/3},
\eeq
where $\O$   is the volume
of the transverse 2-dimensional space.
Since we are working with  dimensionless  coordinates (see the metric (\ref{met2}))
the \bb temperature and  energy are also dimensionless. The
physical dimensions can however easily be restored inverting (\ref{j5}).
Note that the scale of the temperature depends on the normalization
of the time \coo $t$, which is arbitrary.

In the Euclidean action formalism, the thermodynamical potentials are
given as boundary terms of the action at infinity and on the horizon
$r=r_{h}$.
Using the parametrization of the metric,
\beq\lb{par}
ds^{2}= N^2 \Lambda dt^{2}  + \frac{dr^{2}}{\Lambda}  + R^{2}¥dx_{i}dx^{i},
\eeq
the gravitational and scalar
part of the variation of the boundary terms are given respectively by
\cite{Martinez:2004nb}
\bea
\delta I_{G}&=& \frac{2 \Omega}{T} \left[N(RR'\delta \L - \L'R \delta
R)+ 2\L R(N\delta R'- N'\delta R)\right]|^{\infty}_{r_{h}¥},\\
\delta I_{\phi}&=& \frac{4 \Omega}{T} NR^{2}\L \phi' \delta \phi|^{\infty}_{r_{h}¥}.
\eea

The contributions of the two  boundaries  at $r=\infty$ and
$r=r_{h}$ are given by
\bea\lb{bt1}
\delta I_{G}^{\infty}&=& - \frac{2\O}{ T(1+3\b^{2})}\left[\delta
\n_{1}+ \frac{2\b^{2}}{\g}(2- \b^{2})\delta
\n_{2}\right],\nonumber \\
\delta I_{\phi}^{\infty}&=&  \frac{4\O\b^{2}}{\g T(1+3\b^{2})}
\delta \n_{2},\nonumber\\
\delta I_{G}|_{r_{h}}&=&  \frac{2\O}{
T(1+3\b^{2})}\frac{1}{(\n_{1}+\n_{2})}\left[(\n_{1}+\g \n_{2})\delta
\n_{1}+ \b^{2}\n_{1}¥\delta
\n_{2}\right],\nonumber\\
\quad  \delta I_{\phi}|_{r_{h}}&=&0.
\eea

One can easily show that $\delta I_{G}|_{r_{h}}= 4\pi \O \delta
R^{2}(r_{h})= \delta S$. This gives the entropy
\beq\lb{en}
S=I_{G}|_{r_{h}}.
\eeq
On the other hand, because  $\delta I_{\phi}|_{r_{h}}=0$, there is no
thermodynamical potential associated to the scalar field. From
the definition of the the free energy $F=M-TS$ and
from $F=- IT$, it  follows $M=TS - TI= - T(\delta I_{G}^{\infty}+\delta
I_{\phi}^{\infty})$. Using Eqs.\ (\ref{bt1}) one finds
\beq\lb{mass}
M= \frac{2\O}{1+3\b^{2}}\left (\n_{1}+ 2\b^{2}\n_{2}\right).
\eeq

To check the correctness of our calculations, let us consider two
particular cases, $\beta=0$ and $\n_{2}=0$. For $\beta=0$ our solution becomes the
Schwarzschild-AdS black brane and it is easily seen that
Eqs.\ (\ref{TS}) and (\ref{mass}) imply $dM=TdS$. For $\n_{2}=0$ our solution
becomes the usual black brane solution with the $\phi\sim \log r$
singularity at $r=0$ \cite{Cadoni:2011nq}, and again Eqs.\ (\ref{TS}) and (\ref{mass})
entail $dM=TdS$.

Let us now use the constraint (\ref{con1}) to express $M,T,S$  in
terms  of the single parameter $\n_{2}$, with $0\le\n_{2}\le 1$.
We have
\bea\lb{TS1}
T&=&\frac{1}{4\pi}\frac{3\g}{1+3\b^{2}}\n_{2}^{\frac{3\b^{2}-1}
{3\g}}(1-\n_{2}^{2})^{1/3}, \quad S=  4\O\pi
\n_{2}^{-\frac{2}{3\g}}(1-\n_{2}^{2})^{2/3},\\
M&=& \frac{2\O}{1+3\b^{2}}\left
(\frac{1}{\n_{2}}+(2\b^{2}-1)\n_{2}\right),\quad dM= \frac{2\O}{1+3\b^{2}}\left
(-\frac{1}{\n_{2}^{2}¥}+(2\b^{2}-1)\right)d\n_{2}¥,\\
TdS&=&\frac{2\O}{1+3\b^{2}}
\left (\g d\n_{1}+\b^{2}¥\n_{1} d\log(\n_{1}+\n_{2})\right)=
\frac{2\O}{1+3\b^{2}}
\left
(-\frac{1}{\n_{2}^{2}¥}+(2\b^{2}-1)\right)d\n_{2}.
\eea

The first principle of thermodynamics $dM=TdS$ is therefore
satisfied.

\section{Phase transition}
\lb{sect:pt}
In this section we  discuss the thermodynamical properties of our solutions.
The thermodynamical behavior of the \bb
depends crucially on the value of $\b$. It is in particular evident
from Eq. (\ref{TS1}) that $\b^{2}=1/3$  is a transition value.   We
will therefore  discuss separately the three cases $\b^{2}<1/3,\,
\b^{2}=1/3,\, \b^{2}>1/3$.
\subsection {The $\b^{2}<1/3$ case }
For $\b^{2}<1/3$ the scalar \bb exists  in the
full  range  of temperatures, $0\le T<\infty$, with $\n_{2}=1$, 0 corresponding,
respectively to $T= 0$, $\infty$ and to $S=0$, $\infty$.
Notice that the parameter range $\b^{2}<1/3$ corresponds to the
region for which the scalar \bb solution has an holographic
interpretation and the $r=\infty$ region is timelike.

Let us now compute the free energy $F=M-TS$ for the scalar black
brane solution (\ref{met2}) and
for the SAdS solutions (\ref{SADS}). For the scalar \bb we get
\beq\lb{f5}
F_{SB}(T)= \frac{\Omega}{1+3\b^{2}}\left(\frac{3\b^{2}-1}{\n_{2}}+
(\b^{2}+1)\n_{2}\right),\eeq
where $\n_{2}=\n_{2}(T)$ is defined  implicitly by Eq. (\ref{TS1}).
For the SAdS \bb we simply have
\beq\lb{f6}
F_{SAdS}(T)= -\Omega \left(\frac{4\pi}{3}\right)^{3}T^{3}.
\eeq
It is easy to check using Eqs.\ (\ref{f5}) and (\ref{TS1}) that  $F_{SB}(T)$ is
a  monotonic decreasing function of $T$.

One can now  show that  $\Delta  F=F_{SB}-F_{SAdS}$ is positive  for
small $T$, but becomes negative at large $T$.
This can be seen by first considering the small-$T$ ($\n_{2}\sim 1$)
behavior of $F_{SB}$,
\beq\lb{g5}
F_{SB}(T)= \Omega\left( \frac{4\b^{2}}{3\b^{2}+1}-
\frac{2(4\pi)^{3}}{27\g^{2}}(3\b^{2}+1)^{2}T^{3}+{\cal O}(T^{6})\right).
\eeq
Notice that this small-$T$ behavior is dictated by the  $T=0$,
AdS$_{4}$ extremal limit and corresponds to a  holographically  dual 3D
CFT for which the free energy scales as $F\sim T^{3}$.
On the other hand the large-$T$ ($\n_{2}\sim 0$)  behavior is given
by
\beq\lb{g6}
F_{SB}(T)= \Omega\, \frac{3\b^{2}-1}{3\b^{2}+1}\left(
\frac{4\pi
(3\b^{2}+1)}{3\g}\right)^{\frac{3\g}{1-3\b^{2}}}T^{\frac{3\g}{1-3\b^{2}}}.
\eeq
This large-$T$ scaling behavior of the free energy can be also
interpreted
in terms of hyperscaling violation in the dual QFT (see Sect. \ref{sect:hyper}).

The free energy starts positive at small $T$, with $F(T=0)= \Omega\,
\frac{4\b^{2}}{3\b^{2}+1}>0$, so that $\Delta  F>0$,   but,
being $3\g/(1-3\b^{2})>3$,  $\Delta  F$ turns negative at large $T$.
This implies the existence of a critical temperature $T_{c}$ at which
$\Delta  F(T_{c})=0$.
$T_{c}$ can be determined graphically and numerically. By equating
$F_{SB}=F_{SAdS}$ we get :
\beq\lb{k5}
g(y)=1- \b^{2}\frac{3+y}{1-y}=f(y)=
\frac{\g^{3}}{(1+3\b^{2})^{2}}\ y^{\frac{\b^{2}}{\g}},
\eeq
where $y=\n_{2}^{2}$.

One can easily realize  that the two curves $f(y)$ and $g(y)$ do not
intersect for $\b^{2}\ge 1/3$, while they intersect  at a finite non-vanishing value of
the temperature for $\b^{2}<1/3$.

\begin{figure}[ht]
\begin{center}
\begin{tabular}{cc}
\epsfig{file=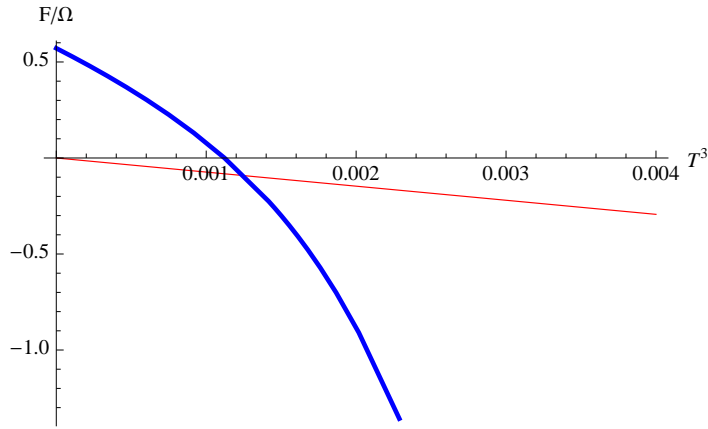,width=9.3cm,angle=0}
\end{tabular}
\caption{Plot of the free energy density $F/\O$ of the scalar \bb for $\b^{2}=1/4$
(blue, thick line) and  of the SAdS \bb (red, thin line) as  function of $T^{3}$.
\label{fig:figure}}
\end{center}
\end{figure}

In  figure \ref{fig:figure} we show
the behavior of the free energy density for  a selected value of $\b$ 
($\b^{2}=1/4$) in the
range $0\le \b^{2}<1/3$. For $\b^{2}=1/4$ we  have  also computed numerically the critical
temperature, that results $T_{c}=0.109583$.

It is also of some interest to determine the behavior of $F$ near
$T_{c}$, i.e.\ its  cross-over  between the two phases. Near $T_{c}$,
 $F$  has $\beta$-independent behavior, and scales as
 \beq\lb{r1}
 F_{SB}\sim\left(\frac{T_{c}-T}{T_{c}}\right)^{\alpha},
 \eeq
with $\alpha=1/n$, where the  integer $n$ is the order of the first non-vanishing
derivative of $T$ evaluated at $T_c$.
For $\b^{2}<1/3$, $(dT/d{\n_{2}})(T_{c})\neq 0$ so that $n=1$.

Summarizing, in the case $\b^{2}<1/3$, above the critical temperature $T_{c}$ the free
energy $F$ of our \bb solution ({\ref{met2}}) becomes smaller than the
free energy of the SAdS solution, and therefore
the SAdS solution becomes unstable and the system
undergoes a first-order phase transition. The phase transition is
first-order because at $T=T_{c}$,
$dF_{SB}/dT\neq dF_{SAdS}/dT$.  The first derivative of free energy and
the specific heat $(dF/dT)+ S$ are therefore discontinuous at $T=T_{c}$.
The presence of this phase transition
has to be understood as a cross-over  from  a  $T<T_{c}$ regime, in
which the  AdS solution  (\ref{ADS}) with vanishing scalar field
is energetically preferred, to  $T>T_{c}$
regime, in which  the scale covariant, \DW solution (\ref{DW}) dressed with scalar
hair is energetically preferred.

The phase transition found  in this section can  also be
described, holographically, in terms of the dual QFT.
The scalar field $\phi$ has to be interpreted in the
dual field theory as an
order parameter, or equivalently a VEV of a scalar operator $\langle
O\rangle$
controlling a phase transition between a phase in which  $\langle
O\rangle=0$ (on the gravity side, the SAdS phase) and a phase in which
$\langle
O\rangle\neq 0$ (on the gravity side, the scalar black brane phase
endowed with a non trivial scalar field). At small temperature  the
behavior of the system is determined by  the IR  fixed point with
$\langle
O\rangle=0$. At large temperatures, the
system is ruled by UV physics, in which the scalar-dressed phase
is energetically preferred and
develops a large negative $\phi$ (corresponding to large negative
values of $\langle O\rangle$),
with a  vanishing self-interaction potential $V(\phi)$.

\subsection {The $\b^{2}=1/3$ case }

For  $\b^{2}=1/3$, it is evident from Eq.\ (\ref{TS1}) that, since
 $0\le \n_{2}\le 1$,  the scalar black  brane solution exists only for
temperatures  below a critical temperature $T=T_{c}=1/4\pi$.
Above $T_{c}$ only the SAdS solution (\ref{SADS}) exists.
This behavior, namely the existence of scalar-dressed solution  only below a critical
temperature,  has been found, numerically, in several
Einstein-Maxwell-scalar models.

For $\b^{2}= 1/3$ the free energy  $F_{SB}(T)$ in Eq.\ (\ref{f5})
can be written explicitly.
We have
\beq\lb{h1}
F_{SB}=\frac{2\O}{3}\sqrt{1- (4\pi T)^{3}}.
\eeq
The free energy is positive definite  and  vanishes for $T=T_{c}$,
whereas $F_{SAdS}$ is always negative. Also in the temperature  range $T\le
T_{c}$, where the scalar \bb exists, we
have $F_{SAdS}<F_{SB}$ and the  SAdS solution is always energetically
favored.
Although this phase is unstable with respect to the SAdS phase,
it is nevertheless interesting to
determine the scaling behavior of $F_{SB}$.

Since for $\b^{2}=1/3$ $(dT/d{\n_{2}})(T_{c})= 0$
but $(d^{2}T/d{\n_{2}^{2}¥})(T_{c})\neq 0$,
near $T_{c}$ the scaling behavior is given by Eq.\ (\ref{r1}) with
$n=2$,
\beq\lb{j6}
F_{SB}\sim \left(\frac{T_{c}-T}{T_{c}}\right)^{1/2}.
\eeq
Near $T=0$, which corresponds to the  AdS$_{4}$ spacetime, we have the
expected scaling behavior of a  CFT in $2+1$ dimensions:
\beq\lb{j7}
F_{SB}= \frac{2\O}{3}\left( 1 -\frac{ (4\pi T)^{3}}{2}\right).
\eeq

\subsection {The $\b^{2}>1/3$ case }
\begin{figure}[ht]
\begin{center}
\begin{tabular}{cc}
\epsfig{file=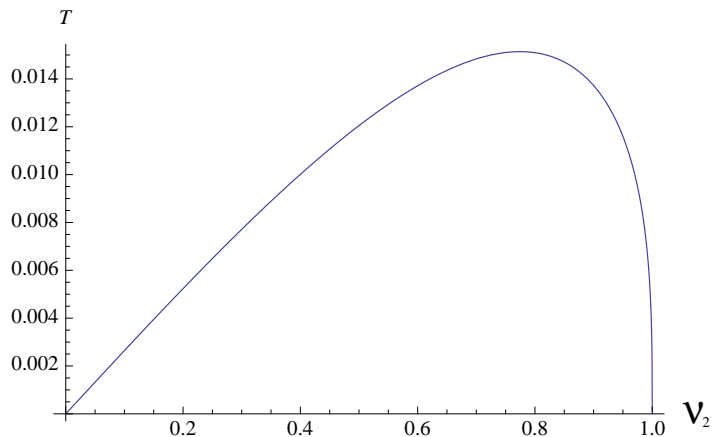,width=9.3cm,angle=0}
\end{tabular}
\caption{Plot of the function $T(\n_{2})$ for $\b^{2}=2/3$.}
\label{fig:figure2}
\end{center}
\end{figure}

\lb{sect:b}
For  $\b^{2}>1/3$ the function  $T(\nu_{2})$ of Eq.\ (\ref{TS1}) is
not monotonic, but in the range of definition of $\n_{2},\,$
$0\le\nu_{2}\le 1$ it has a local maximum
at $\n_{2}= \n_{0}= \sqrt{(3\b^{2}-1)/(\b^{2}+1)}$.
This means that also in this case the scalar \bb solution exists only
below a critical temperature $T_{c}\id T(\n_{0})= 3\g^{4/3}
[2^{5/3}(3\b^{2}-1)^{(1-3\b^{2})/6\g}¥\p(1+3\b^2)(1+\b^2)^{(1+\b^2)/6\g}]^\mo$.
Notice that $F$ has a minimum in $\n_0$.
In figure \ref{fig:figure2} we show the function $T(\n_{2})$ for
$\b^{2}=2/3$. In this case $\n_{0}=\sqrt{3/5}$ and
$T_{c}=2^{1/3}5^{-5/6}¥
/(\sqrt{3}\, 4\pi)$.

\begin{figure}[ht]
\begin{center}
\begin{tabular}{cc}
\epsfig{file=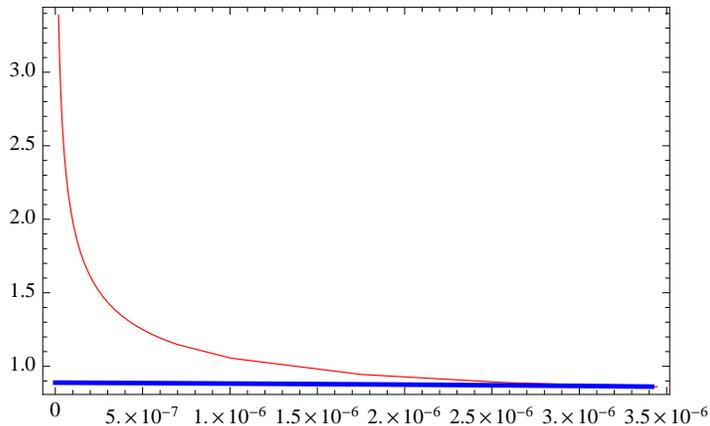,width=9.3cm,angle=0}
\end{tabular}
\caption{Plot of the free energy density $F_{SB}/\O$ as a function of
$T^{3}$ for the two branches of
the solution   in the $\b^{2}=2/3$ case. The blue (thick) line
indicates  the first (AdS$_{4}$) branch, whereas the red (thin) line
indicates the second branch.}
\label{fig:figure3}
\end{center}
\end{figure}

For $T\le T_{c}$, our model admits both the scalar \bb and the SAdS
solutions. Above $T_{c}$, only the SAdS solution exists.
Differently from the $\b^{2}=1/3$ case,  the non-monotonicity of
$T(\n_{2})$ implies the existence of two different branches of the
scalar-dressed  \bb phase  for $T\le
T_{c}$.
From Eq.\ (\ref{f5}) one easily realizes that  for $\b^{2}>1/3,\,
F_{SB}$  is always  positive, and hence
$F_{SB}>F_{SAdS}$, implying instability of the two
phases  with respect to the SAdS phase.
The behavior of the free energy density for the two branches is
shown in Fig.\ \ref{fig:figure3} for  $\b^{2}=2/3$.

Even if they are unstable, it is
however of interest to briefly discuss the features of  these small-$T$ phases.
The first branch is obtained for  $\n_{0}\le \n_{2}<1$ and
is the usual AdS$_{4}$ phase we  have already obtained at small $T$
when $\b^{2}\le 1/3$. As  expected, in this case the free energy
scales for $T\sim T_{c}$ and $T\sim 0$ as in Eqs.\ (\ref{j6}) and
(\ref{j7}), respectively.

The second branch is obtained for $0\le\n_{2}\le \n_{0}$ and has no
analogue  for  $\b^{2}\le 1/3$.  The free energy scales at small
temperatures as in Eq. (\ref{g6}),  $F\sim T^{\alpha},\,
\alpha=3\g/(1-3\b^{2})$. But now $\alpha$ is negative  and we have a
singularity of $F_{SB}$ at $T=0$. This small-$T$ singular
behavior  is due to the fact that  for $\n_{2}=0$ there is no $T=0$
extremal soliton, but an extremal domain  wall solution with the
scalar behaving as  $\log r$.

Near $T_{c}$ the scaling behavior of $F_{SB}$ is the same as in the
$\b^{2}=1/3$ case. It is given by Eq. (\ref{j6}).

\section{Hyperscaling Violation}
\lb{sect:hyper}

The  different thermodynamical phases    of our  scalar \bb
and their peculiar behavior  described in the previous sections
can also be understood in terms of its symmetries
in the UV and IR regimes.

In particular, this thermodynamical
pattern finds its natural explanation in terms of the UV hyperscaling violation
generated in the dual QFT by the asymptotic  DW solution (\ref{DW}) with respect to the
IR scaling-preserving solution given by the extremal AdS$_{4}$  spacetime
(\ref{ADS}). In fact, the hyperscaling-violating phase is
associated with the emergence of an UV length scale $r_{0}$, which
obviously  decouples in the IR conformal phase.

This behavior is peculiar and somehow unexpected, because  the  
regime of the dual QFT in
which one naturally expects hyperscaling violation to occur is not the UV but
the IR. This is not only true for the condensed matter systems for
which hyperscaling violation was originally discovered \cite{Fisher:1986zz}, but
also for   its   recent holographic realizations \cite{Huijse:2011ef,Dong:2012se,Kim:2012nb}.

The description of holographic  hyperscaling violation in $d+2$
dimensions  is based on
the scale covariant  metric \cite{Huijse:2011ef}\footnote{In the literature there are
several, equivalent, definitions  of scale covariant metrics
generating in the dual QFT  hyperscaling violation.
They are related  by reparametrizations of the radial coordinate $r$.
Here, we use the  metric of reference \cite{Huijse:2011ef}, but the same
results can be obtained using for instance the metric of reference 
\cite{Dong:2012se}.}
\beq\lb{hv}
ds^{2}= \frac{1}{r^{2}}\left(- \frac{dt^{2}}{r^{2d(z-1)/(d-\theta)}}
+\OM+ r^{2\theta/(d-\theta)}dr^{2}\right),
\eeq
where $d$ is the number of transverse dimensions,
$\theta$ is the hyperscaling violation parameter and $z$ is the
dynamic critical exponent (it describes
anisotropic scaling, hence  violation of Poincar\'e  symmetry, in the
$d+1$ spacetime). The transformation law  under rescaling of the
coordinates is
\beq\lb{tl}
t\to \l^{z} t,\quad x_{i}\to \l x_{i},\quad r\to \l^{(d-\theta)/d}
r,\quad ds\to \l^{\theta/d} ds.
\eeq

For $\theta=0$ the metric (\ref{hv}) is scale invariant, hence for
generic values of $z$ it gives the Lifshitz spacetime. For $z=1$ the
$d+1$-dimensional
spacetime has  Poincar\'e isometry. Finally,  for $\theta=0$, $z=1$,
the metric (\ref{hv}) becomes the  AdS$_{d+2}$ spacetime with both Poincar\'e
isometry in the $d+1$
sections and scaling (conformal) symmetry. Taking into account that
the the temperature scales as  an inverse time, in presence of a
black brane  the area law  together with Eq.\ (\ref{tl}) implies the
following scaling behavior for the free energy:
\beq\lb{scal}
F\sim T^{\frac{(d-\theta)+ z}{z}}.
\eeq
This relation allows a simple physical interpretation of the
hyperscaling violation parameter $\theta$, which is analogous to that
used in condensed matter critical system: the hyperscaling relation
between specific heat $\hat \alpha$  and critical exponent $\hat\nu$, given by
$2-\hat \alpha=d\,\hat\nu$  is modified by ``lowering'' the dimensionality of
the system from  $d$ to $d-\theta$, namely
$2-\hat\alpha=(d-\theta)\hat\nu$.

The $r=\infty$ asymptotic form of our scalar \bb solutions (\ref{DW})
(we set here $r_{0}=1$)
can be brought in the form (\ref{hv}) by a redefinition of the radial
coordinate,
\beq\lb{rd}
r\to r^{-(1+3\b^{2})}.
\eeq
Notice that using the radial coordinate of the metric  (\ref{hv}) the UV
region of the dual QFT is described by $r=0$, whereas the IR region
corresponds to $r=\infty$. Consistently,  with Eq. (\ref{rd})  the
role played by $r=0$ and $r=\infty$ is reversed when one uses the
radial coordinate of the metric (\ref{DW}).

Taking into account that in our case $d=2$, we easily find that the
asymptotic \bb metric (\ref{DW}) is characterized by  the following
parameters:
\beq\lb{para}
z=1,\quad \theta= \frac{6\b^{2}}{3\b^{2}-1}.
\eeq
On the other hand the values of the parameters $z$ and $\theta$ for the
extremal AdS$_{4}$ solution (\ref{ADS}) are obviously given by
\beq\lb{para1}
z=1,\quad \theta=0.
\eeq

The value $z=1$, implying Poincar\'e isometry in the
$(d+1)$-dimensional  spacetime, is largely expected for a scalar,
electromagnetically neutral, solution. In fact it is
well known that in order to obtain $z\neq 1$ (Lifshitz) solutions, the
brane must carry $U(1)$ charge. Inserting in Eq.\  (\ref{scal}) $z=1$,
$d=2$, we get
\beq\lb{k7}
F\sim T^{3-\theta}.
\eeq
As a check of the correctness of our result, one can recover Eqs.\
(\ref{g6}) and (\ref{g5}) inserting, respectively, Eqs.\ (\ref{para})
and (\ref{para1}) into Eq.\ (\ref{k7}).

As  expected, we have $\theta\neq 0$ in the  scalar \bb phase. This
gives the deviation from the conformal scaling
of the free energy of  a $2+1$ conformal field theory, which one
obtains  for $\theta=0$, in the hyperscaling-preserving phase.

On the other hand, the particular values (\ref{para})
of the hyperscaling-violation parameter for our model
are rather intriguing. The parameter
$\theta$ is {\sl negative} for $\b^{2}<1/3$,  diverges for
$\b^{2}=1/3$ and becomes positive, with $\theta\ge 3$, for
$\b^{2}>1/3$. Although $\theta$ is negative, one can easily realize
that  the null energy
conditions for the bulk stress-energy tensor  are satisfied.  In fact
for $z=1$ these conditions require either $\theta\le 0$ or $\theta\ge
d$ \cite{Dong:2012se}.

The negative value of $\theta$ for $\b^{2}<1/3$ has no
counterpart in  condensed matter critical system, for which
the hyperscaling-violating parameter is positive. This is probably
related to another striking difference between the two  cases.
In the condensed matter case the hyperscaling-violating phase is
stable at
small temperatures where (for instance   random-field induced) long-scale
fluctuations dominate over thermal fluctuation.  In our gravitational--holographic
case the opposite happens. The hyperscaling-violating
phase  is  stable   at large temperatures, but becomes unstable at
small temperatures, where the hyperscaling-preserving phase is
energetically preferred.
This interchange of IR and UV physics is a rather puzzling point,
 whose physical meaning is presently not clear
to us. It is possible that  negative values of $\theta$ are related to
the fact that our gravitational system
at large temperatures   seems  to prefer to live in   more than 3+1
dimensions. This possibility  is also supported by the fact that our 
model arises from compactifications of black $(p-1)$-branes \cite{Gouteraux:2011ce}.
Moreover, negative
values of $\theta$ have been obtained in  some string theory
constructions based on  Dp branes \cite{Dong:2012se} 

Naively, one could hope that the behavior  in the $\b^{2}>1/3$ case would
shed light on these puzzling features. Although the hyperscaling-violating phase
(the second branch discussed in Subsection \ref{sect:b})
 is in this case unstable, nevertheless it occurs at small
temperatures, is characterized by a positive $\theta$ and the null 
energy condition $\theta>d$ \cite{Dong:2012se} for the bulk 
stress-energy tensor is 
satisfied.
Unfortunately, in this case  $\theta$ is greater than 3, so that the free
energy (\ref{k7}) scales at small temperatures with a negative
exponent (see also Eq.\ (\ref{g6})) and becomes singular at $T=0$.

\section{ The $\b^{2}>1$ case}
\lb{sect:bmuno}
The solutions for $\b^2>1$ have also been obtained in
\cite{Cadoni:2011yj}. Actually, due to the
invariance of the action for $\b\to1/\b$, they can simply be recovered from (\ref{met1})
by duality, substituting everywhere $\b$ with $1/\b$. Note that for $\b^2>1$ the
parameter $\g$ becomes negative.

The solutions can be written as

\bea\lb{met3}
ds^2=&&\Delta(r)^{-\frac{2}{3\g}}\left[\left(\frac{r}{r_0}\right)^{\frac{2\b^{2}¥}
{3+\b^2}}\left( -\Gamma(r) dt^2+\OM\right)+
E\,\frac{\Delta(r)^{-\frac{4}{3\g}}}{\Gamma(r)}\left(\frac{r}{r_{0}}\right)^{-\frac{2\b^{2}¥}
{3+\b^2}}\,dr^2 \right],\cr
e^{2\f}=&&\left[\frac{\Delta(r)}{D}\left(\frac{r}{r_0}\right)
^{-\frac{3\g}{3+\b^2}}\right]^{\frac{2\b}{\sqrt3\g}},
\eea
where
$\Gamma(r)= 1-{\m_1}\left(\frac{r_0}{r}\right)^\d$,
$\Delta(r)= 1+{\m_2}\left(\frac{r_0}{r}\right)^\d$, $\d=-3\g/(3+\b^{2})\,$,
$D= \sqrt{\m_{2}(\m_{1}+\m_{2})}$,
$E=\left(\frac{\g L}{(3+\b^{2})r_0}\right)^{2}D^{2/\g}$,
and $\m_1\ge 0,\m_{2}\ge 0$ are free parameters.

For $r\to\infty$, the solutions behave as domain walls
\beq\lb{DW2}
ds^{2}= r^{\frac{2\b^2}{3+\b^2}}\left( -dt^{2}+\OM\right)+r^{-\frac{2\b^2}{3+\b^2}}dr^{2}.
\eeq
with $\phi= -[(\sqrt{3}\b)/(3+\b^{2})]\log r$.

The properties of the the metric (\ref{met3}) are analogous to those holding
in the $\b^2<1$ case. If $\m_2\ge 0$, it exhibits a singularity at $r=0$,
shielded by a horizon at $\frac{r}{r_0}=\m_1^{\,1/\d}$, and therefore represents a
regular black brane.

For  $\m_{2}>0$, the extremal limit $\m_1=0$,  of solution (\ref{met3})
represents a regular soliton with the same behavior as that
described in the $\b^{2}<1$ case.

The solution can be written in a simpler form by rescaling the
coordinates and defining new parameters $\n_1$ and $\n_2$, as
in the $\b^2<1$ case:
\bea\lb{met4}
ds^2=&&\left(\frac{\g L}{3+\b^{2}}\right)^{2}\left[- \Delta(r)^{- \frac{2}
{3\g}}\left(1-\frac{\n_1}{r^\d}\right)r^{\frac{2\b^2}{3+\b^2}}dt^2+
\frac{ \Delta(r)^{-\frac{2}{\g}}\,dr^2}{(1-\frac{\n_1}{r^{\d}}) r^{\frac{2\b^2}{3+\b^2}}}\right]
+ \Delta(r)^{-\frac{2}{3\g}}r^{\frac{2\b^2}{3+\b^2}}\OM,\cr
&&\cr
e^{2\f}=&&\Delta(r)^{\frac{2\b}{\sqrt3\,\g}}r^{-\frac{2\sqrt3\,\b}{3+\b^2}}.
\eea
The new parameters are not independent, but satisfy the relation (\ref{con1}).
Thus, in this form the solution has only one independent
parameter, $\n_{2}$, with $0<\n_{2}\le 1$.

The thermodynamics of the $\b^2>1$ solutions can be obtained repeating the
calculations of sect.\ \ref{ther}, or simply by duality from the case $\b^2<1$.

For the temperature $T$ and entropy $S$  of
the solution (\ref{met4}) we get
\beq\lb{TS2}
T=-\frac{1}{4\pi}\frac{3\g}{3+\b^{2}}(\n_{1}+\n_{2})^{\frac{2}
{3\g}}\n_{1}^{1/3}, \qquad S=  4\pi \O
(\n_{1}+\n_{2})^{-\frac{2}{3\g}}\n_{1}^{2/3},
\eeq
while the mass is
\beq\lb{mass2}
M= \frac{2\O}{3+\b^{2}}\left (\b^2\n_{1}+ 2\n_{2}\right).
\eeq

We can now use the constraint (\ref{con1}) to express $M,T,S$  in
terms  of the independent parameter $\n_{2}$. They read
\beq\lb{MTS2}
M= \frac{2\O}{3+\b^{2}}\left
(\frac{\b^2}{\n_{2}}+(2-\b^{2})\n_{2}\right),\quad
T=\frac{1}{4\pi}\frac{-3\g}{3+\b^{2}}\n_{2}^{\frac{\b^{2}-3}
{3\g}}(1-\n_{2}^{2})^{1/3}, \quad S=  4\O\pi
\n_{2}^{\frac{2\b^2}{3\g}}(1-\n_{2}^{2})^{2/3},
\eeq
It is easy to check that the first principle of thermodynamics $dM=TdS$ is
satisfied.

Finally, the free energy $F=M-TS$ for the black brane solution (\ref{met4})
is given by
\beq\lb{g4}
F_{SB}(T)= \frac{\Omega}{3+\b^{2}}\left(\frac{3-\b^{2}}{\n_{2}}+
(1+\b^{2})\n_{2}\right),\eeq
where $\n_{2}=\n_{2}(T)$ is defined  implicitly by Eq. (\ref{TS2}).

The thermodynamical properties of the solutions with $\b^2>1$ follow
from the previous discussion of the $\b^2<1$ case by duality.
In particular,  one must distinguish three cases,
$\b^{2}>3,\, \b^{2}=3,\, \b^{2}<3$. We summarize their properties.
\subsection {The $\b^{2}>3$ case }
This case is analogous to $\b^2<1/3$.
For $\b^{2}>3$, the temperature of the \bb can range from 0 to $\inf$.
It is easy to check using Eqs.\ (\ref{TS2}) and (\ref{g4}) that
$F_{SB}(T)$ is a  monotonic decreasing function of $T$.
Moreover, $\Delta  F=F_{SB}-F_{SAdS}$ is positive  for
small $T$ but becomes negative at large $T$.
In fact, the small-$T$ behavior of $F_{SB}$  is
\beq\lb{h5}
F_{SB}(T)= \Omega\left( \frac{4}{3+\b^{2}}-\frac{(4\pi)^{3}}
{27\g^{2}}(3+\b^{2})^{2}T^{3}+{\cal O}(T^{6})\right),
\eeq
while the large-$T$ behavior is given by
\beq\lb{h6}
F_{SB}(T)= \Omega \frac{3-\b^{2}}{3+\b^{2}}\left(\frac{4\pi
(\b^{2}+3)}{-3\g}\right)^{-\frac{3\g}{\b^{2}-3}}T^{-\frac{3\g}{\b^{2}-3}}.
\eeq

Hence, $\D F$ is positive for $T$ small, but becomes negative at large $T$.
This implies the existence of a critical temperature $T_{c}$ at which
$\Delta  F(T_{c})=0$.

Therefore, when $\b^{2}>3$,  the free energy $F$ of the \bb solution
(\ref{met4}) becomes smaller than the
free energy of the Schwarzschild-anti de Sitter (SAdS) solution
above the critical temperature $T_{c}$, and hence the SAdS solution
becomes unstable and the system undergoes a first-order phase transition.

Analogously to  $\b^{2}<1$ also in the $\b^{2}>1$ case,  we can
characterize the solutions with the dynamic critical  exponent $z$ and the
hyperscaling-violation parameter $\theta$. We have
\beq\lb{l1}
z=1,\quad \theta=\frac{6}{3-\b^{2}}.
\eeq
\subsection {The $\b^{2}=3$ case }

It is evident from Eq.\ (\ref{TS2}) that for  $\b^{2}=3$, the temperature
of the black  brane solution is
always lower than a critical temperature $T_{c}=1/4\pi$.
Above $T_{c}$ only the SAdS solution (\ref{SADS}) exists.

For $\b^2=3$, the free energy  $F_{SB}(T)$ can be written explicitly,
and is equal to that of the $\b^2=1/3$ \bb (\ref{h1}).
The thermodynamics of the $\b^2=3$ \bb is therefore identical to that
discussed in the $\b^2=1/3$ case.

\subsection {The $\b^{2}<3$ case }
As for $1/3<\b^2<1$, also for $1<\b^{2}<3$ the function  $T(\nu_{2})$
is not monotonic, but has a local maximum, located at
$\n_{0}= \sqrt{(3-\b^{2})/(1+\b^{2})}$.
Therefore, also in this case the temperature of the scalar \bb solution
must be less than a critical temperature
$T_{c}\id T(\n_{0})= 3\g^{4/3}(3-\b^2)^{(\b^2-3)/6\g}
(1+\b^2)^{(1+\b^2)/6\g}[2^{5/3}\p(3+\b^2)]^\mo$.

For $T\le T_{c}$ our model admits both the scalar \bb and the SAdS
solutions. Above $T_{c}$ only the SAdS solution exists.
The non-monotonicity of
$T(\n_{2})$ implies the existence of two different branches of the
scalar-dressed  \bb phase  for $T\le
T_{c}$.
Since in the present case $F_{SB}$  is always  positive,  we
necessarily  have $F_{SB}>F_{SAdS}$, implying instability of the two
phases  with respect to the SAdS phase. 

The first branch is obtained for  $\n_{0}\le \n_{2}<1$ and
is the usual AdS$_{4}$ phase. As  expected, the free energy
scales for $T\sim T_{c}$ and $T\sim 0$ as in Eqs.\ (\ref{h5}) and
(\ref{h6}), respectively.
The second branch is obtained for $0\le\n_{2}\le \n_{0}$.
In this case, the free energy scales at small
temperatures as in Eq.\ (\ref{h6}),  $F\sim T^{\alpha},\,
\alpha=3\g/(3-\b^{2})$, but now $\alpha$ is negative  and we have a
singularity of $F_{SB}¥$ at $T=0$.

\section{General models and charged solutions}
\label{sect:gm}

Until now we have restricted our investigation to the uncharged case
and the particular  model described by the potential (\ref{k3}).
In this section we will extend our discussion to  general models
described by Eq. (\ref{b1}) and
to charged solutions.

\subsection{General models}
\lb{sect:gm1}
The main features of the \bb solutions described in the previous
sections are not a peculiarity of the model (\ref{k3}) but
are determined by the  behavior of the potential (\ref{k3}) at $\phi=0$
and $\phi=-\infty$ (see Eqs.\ (\ref{b1})).

In  Ref.\ \cite{Cadoni:2011nq} has been derived   the  general
asymptotic solution  of a model with an exponential potential,
\bea
V(\phi)&=&-\frac{2(3-h^{2})}{(1+h^{2})^{2}}\ e^{2h\phi},\quad
\phi=-\frac{h}{h^{2}+1}\,\log r,\nonumber\\
ds^{2}&=& r^{\frac{2}{h^{2}+1}}\left(-dt^{2}+\OM\right)
+r^{-\frac{2}{h^{2}+1}}dr^{2}.\lb{s1}
\eea

The case described by Eq. (\ref{b1}) is covered by  setting $ h^{2}<
3$ in Eq. (\ref{s1}), so that the $r=\infty$ region corresponds to $\phi=
-\infty$ where the potential approaches zero exponentially.
On the other hand the condition (\ref{b1}) at $\phi=0$ implies the
existence of an AdS$_{4}$ vacuum for $\phi=0$.

Obviously, a generic model will not be exactly
integrable. The existence of a scalar black brane solution
interpolating between the AdS$_{4}$ vacuum at $r=0$ and the scale
covariant
solution (\ref{s1}) at $r=\infty$ can only be proved numerically.
Nevertheless, if such a solution exists, then necessarily the
thermodynamical system for $h^{2}<1$  must undergo the scalar \bb
$\to$ SAdS phase
transition described in Sect.\ \ref{sect:pt}.

This can be shown by first realizing that at small $T$ the free
energy of the scalar \bb must have a behavior similar to that of Eq.\
(\ref{g5}), i.e.\ $F_{SB}= C_{1}- C_{2}T^{3}$, with $C_{1,2}$ positive
constants.
This implies that at small $T$, $F_{SB}-F_{SAdS}>0$.
On the other hand at large $T$ the entropy of a scalar \bb with the
asymptotic behavior (\ref{s1}) scales as $S\sim T^{2/(1-h^{2})}$,
implying the scaling for the free energy $F_{SB}\sim
-T^{(3-h^{2})/(1-h^{2})}$. For $h^{2}<1$ we have
$T^{(3-h^{2})/(1-h^{2})}>T^{3}$, from which follows that at large
$T$,
$F_{SB}-F_{SAdS}<0$. This proves the existence of a scalar \bb
$\to$ SAdS
phase transition and of an associated critical temperature.

Comparing equation (\ref{s1}) with Eq.\ (\ref{hv}) one can read off,
using a suitable reparametrization of the radial coordinate,
the hyperscaling violation  parameter $\theta=2h^{2}/(h^{2}-1)$, and
the dynamic critical exponent $z=1$. Notice that $\theta$ is negative 
for $h^{2}<1$.

Comparison of  the large-$T$ behavior of our models with that characterizing the small-$T$
regime of the usual  models for holographic hyperscaling violation sheds some light on
the nature of the phase transition.
The latter models use  potentials $V(\phi)$ that at $\phi=0$ have a
maximum  instead of a minimum. This corresponds to  a local tachyonic
scalar excitation  near $\phi=0$ with mass slightly above the BF bound.
For $\phi\to\infty$, the potential diverges exponentially. This
$\phi\to\infty$ regime is also described by Eq. (\ref{s1}) with
$h^{2}<3$, but now $\phi\to \infty$ corresponds to the IR region $r=0$,
whereas $\phi=0$ describes the $r=\infty$ UV region.
In some sense the phase transition described in this paper is an UV
counterpart of the  IR phase transition of usual hyperscaling-violating
models.

\subsection{Charged solution}
Let us now briefly discuss  the electrically charged case.
A Maxwell field can be introduced in the action
(\ref{action})
in two different ways, that is with minimal or non-minimal coupling
to the scalar field.
In the minimal case the action  (\ref{action}) becomes
\beq\lb{action1}
I=\int d^{4}x \sqrt{-g}\left[R-2 (\partial \phi)^{2}
-F^{2}-V(\phi)\right].
\eeq
 Let us briefly discuss a model   with a potential $V$ given by Eq.\ (\ref{k3}).
 Needless to say, our considerations can be trivially extended to a
 generic potential satisfying the conditions (\ref{b1}).

For $\phi=$ const.\ $=0$, the  model admits the Reissner-Nordstrom-anti
de Sitter (RNAdS) solution.
For non-trivial scalar field configurations,
the exact integrability valid in the neutral case is lost,  but one can
treat the problem  using the  same method described in the previous
sections.

In the asymptotic region $r\to\infty, \phi\to-\infty$  the potential
behaves exponentially
\beq\lb{s4}
V(\phi)=-\frac{6\b^{2}}{\g L}e^{2\sqrt{3}\b\phi}.
\eeq

In the charged case the asymptotic form of the solution is \cite{Cadoni:2011nq},

\beq\lb{s5}
ds^{2}= - r^{\frac{8-6\b^{2}}{4+3\b^{2}}}dt^{2}+
r^{-\frac{8-6\b^{2}}{4+3\b^{2}}}dr^{2}+ Q r^{\frac{6\b^{2}}{4+3\b^{2}}}
\OM,\qquad
\phi=-\frac{2\sqrt{3}\b}{6\b^{2}+4}\log r,
\eeq
where $Q$ is the electric charge.
In the electrically charged case, the small-$T$, small-$r$ behavior is more
involved than in the uncharged case. This is mainly because in
presence of the electric charge  beside the AdS$_{4}$ vacuum also
AdS$_{2}\times R^{2}$ vacua are possible.

We will not enter into the details, but will  just assume the
existence of
scalar, electrically charged  BB solution with  asymptotic behavior
given by Eq.\ (\ref{s5}), which interpolates between the $T=0$ vacuum
and the asymptotic solution (\ref{s5}). Typically the existence of
this solution has to be shown numerically.
The asymptotic
solution (\ref{s5}) can be put in the form (\ref{hv}) by the
coordinate transformation $r\to r^{-(4+3\b^{2})/3\b^{2}}$.
The parameters $z,\theta$ can be easily calculated. We have
$\theta=4$, $z=3-4/(3\b^{2})$. Notice that in the electrically charged
case the hyperscaling  violation parameter is 
$\beta$-independent, whereas it is the dynamic critical exponent  $z$ that becomes
$\b$-dependent.
Moreover,  the constraints coming from the  null energy condition
for
the bulk stress-energy tensor
(see Ref.\ \cite{Dong:2012se})  are satisfied for $\b^2\le 2/3$.

With these values of $z$ and $\theta$ Eq.\ (\ref{scal}) gives the
large-$T$ behavior of the free energy
\beq\lb{l1a}
F_{SB}\sim - T^{1-\frac{2}{z}}=- T^{\frac{3\b^{2}-4}{9\b^{2}-4}}.
\eeq
The scaling exponent for $F$ becomes greater than $3$ for $\b^{2}>1/3$.
It follows that at large $T$ the free energy of the scalar BB becomes
smaller than the free energy of the RNAdS solution.
At large $T$ a phase transition scalar BB $\to$ RNAdS solution becomes
possible. Obviously, the phase diagram of our solution at small $T$
could be rather complicated. The existence of the phase transition has
therefore to be confirmed by explicit numerical calculations.

To conclude, let us briefly comment on  the case of non-minimal couplings
between the Maxwell and the scalar field. The action is now,

\beq\lb{action3}
I=\int d^{4}x \sqrt{-g}\left[R-2 (\partial \phi)^{2}
-f(\phi)F^{2}-V(\phi)\right],
\eeq
where $f(\phi)$ is chosen  as a combination of exponentials.
These  models are of particular relevance because they
represent the framework in which  holographic hyperscaling violation
first emerged \cite{Goldstein:2009cv,Charmousis:2009xr,Cadoni:2009xm,
Charmousis:2010zz,Li:2012ib,Gouteraux:2011qh}.

Depending on the properties  of the potential $V$ and of the coupling
function $f(\phi)$ the model can  have an hyperscaling violating phase
either in the IR or in the UV.
When the potential has a maximum at  $\phi=0$, the scalar field
$\phi$ behaves as $\phi\sim - \log r$ for $r=0$, in the IR region
we have $\phi\to \infty$ and  the coupling function  behaves as
$f(\phi)\sim e^{\delta \phi}$ we get
hyperscaling  violation in the IR  \cite{Goldstein:2009cv,Charmousis:2009xr,Cadoni:2009xm,
Charmousis:2010zz,Li:2012ib,Gouteraux:2011qh,Dong:2012se}.
Conversely, when the potential has a maximum at  $\phi=0$ (like,
e.g in Eq.\ (\ref{k3})),
$\phi$ behaves as $\phi\sim - \log r$ for $r=\infty$, in the UV region
we have $\phi\to -\infty$ and the coupling function  behaves as
$f(\phi)\sim e^{\delta \phi}$ we get the
hyperscaling  violation in the UV discussed in this paper.

\section{Conclusions}

In this paper we have investigated the thermodynamical behavior and
the scaling symmetries of the BB solutions  of AdS Einstein-scalar
gravity  models for which the squared mass for the scalar is
positive and the potential vanishes exponentially as $\phi\to- \infty$.

Our investigation has been mainly focused on an integrable model, 
which also arises as compactification of black $(p-1)$-branes,
for which exact analytic solutions  can be found. However, we have
been able to show that the
relevant features of this model can easily be extend to a broad class
of Einstein-scalar and Einstein-Maxwell-scalar gravity models.

We have found that this  broad class of models  has an  interesting
thermodynamical phase
diagram  and extremely non-trivial behavior in the
ultraviolet regime of the holographically dual QFT. Hyperscaling violation generates  an UV
length scale, related to the size of the brane, which decouples in the
IR where conformal invariance is
restored. At high temperatures the SAdS solution becomes unstable and
a scalar-dressed BB solution, with non-AdS, scale-covariant asymptotical behavior,
becomes energetically preferred. This new, scalar-dressed phase can
be characterized  by the two   parameters normally used for critical
systems with hyperscaling violation: the dynamical critical exponent
$z$  and the hyperscaling violation parameter $\theta$.

The actual value of the parameters $z$ and $\theta$ depends on the
particular model under consideration, for instance $z=1$ for
electrically uncharged BB and $z\neq 1$ for electrically charged BB.
Moreover,  quite generically the  values of  $z,\theta$ satisfies the constraints
coming from
the null energy conditions \cite{Dong:2012se}.

On the other hand the most striking  feature of our models is that  
for uncharged \bb the
hyperscaling parameter  $\theta$ is always {\sl negative}.

This is a rather puzzling feature, whose full meaning
is presently not  clear  to us. To our knowledge negative values
of $\theta$ are not  realized in usual condensed matter critical
systems. In the holographic framework,  negative values of $\theta$
have been obtained in some string theory
constructions based on  Dp branes \cite{Dong:2012se}.

In the usual scenario for holographic hyperscaling violation, when the
hyperscaling violation phase is stable in the IR, $\theta$ is positive.
This implies that  the scaling law for the free energy is that pertinent
to a CFT in  $d-\theta$ dimensions. This lowering of the
``effective dimensions''  is a crucial ingredient  in the
small temperature behavior of traditional (e.g.\ Ising models)
hyperscaling-violating critical systems \cite{Fisher:1986zz}. The same effect  is also
a key concept in the holographic
explanation of phenomena like hidden Fermi surfaces or the area  law
violation for entanglement entropy \cite{Huijse:2011ef,Dong:2012se,
Shaghoulian:2011aa,Ogawa:2011bz}.
Conversely, the gravitational models investigated in this paper have
a negative hyperscaling-violation parameter  $\theta$ and are therefore
characterized by a raising  of the
``effective dimensions''. For this reason it does not seem  that they
could be relevant   for applications to condensed  matter systems.
It is more likely that they could be useful for understanding the
holographic features of the gravitational interaction.

A possible way to shed light on the UV behavior of our model is
to calculate the correlation functions for the scalar operators in
the dual field theory. For positive  $\theta$ these correlation functions
have been calculated in
Ref.\ \cite{Dong:2012se}.  They are characterized  by an intriguing cross-over behavior
between an exponential form at large distance and an universal,
power-law form at short distances.  Unfortunately, the short-distance calculations
of Ref.\ \cite{Dong:2012se}  do not  hold for negative $\theta$, which is the relevant case
for our models.

Apart from its  interesting ultraviolet features,  our model could  also be  useful
for resolving IR singularities of  scale covariant metrics.
In fact it is known that this kind of metrics are singular at small
radius \cite{Kachru:2008yh,Horowitz:2011gh,Harrison:2012vy}.
The restoration of the conformal invariance and the emergence of AdS
geometry near  $r=0$   realized by our scalar soliton (\ref{met1a}) represent  a
natural  way to cure  the small radius singularities of metrics of
the form (\ref{DW}).

In this paper we have considered  $4$-dimensional Einstein-scalar 
gravity models, however we 
expect that our results can be easily extended to  generic spacetime dimensions.
In particular, we expect that the higher-dimensional generalization 
of our models and solutions will match the general expressions  for compactified black 
$(p-1)$-branes with a smeared charge supported by a $G_{(p+1)}$ field-strength form 
obtained in Ref. \cite{Gouteraux:2011ce}.

\acknowledgements
We thank  Blaise Gout\'eraux
and Elias Kiritsis  for having pointed out to us that the model and 
the solutions discussed in this paper also arise as compactification 
of black $(p-1)$-branes with a smeared charge.


\bibliography{dw1}

\end{document}